\documentclass{article}
\usepackage{arxiv}
\usepackage{caption}
\usepackage{epsfig}
\usepackage{amssymb}

\usepackage{supertabular}
\usepackage{array}

\usepackage{color}

\usepackage{mathtools}

\usepackage[utf8]{inputenc} 
\usepackage[T1]{fontenc}    
\usepackage{hyperref}       
\usepackage{url}            
\usepackage{booktabs}       
\usepackage{amsfonts}       
\usepackage{nicefrac}       
\usepackage{microtype}      
\usepackage{lipsum}
\usepackage{graphicx}
\usepackage{multirow}

\title{A full electromagnetic Particle-In-Cell code to model collisionless plasmas in magnetic traps} 

\author{	
E. A. Orozco \\ Universidad Industrial de Santander \\ Bucaramanga, Colombia \\ 
	\And 
V. D. Dugar-Zhabon \\ Universidad Industrial de Santander \\ Bucaramanga, Colombia \\
     \And 
Alex Estupi\~n\'an\thanks{\textit{E-mail address:} alex.estupinan@saber.uis.edu.co} \\ Universidad Industrial de Santander \\ Bucaramanga, Colombia \\ 
      \And 
M. T. Murillo Acevedo\\ Universidad Manuela Beltr\'an \\ Bucaramanga, Colombia \\
}

\begin{document}
\maketitle

\begin{abstract}
A lot of plasma physics problems are not amenable to exact solutions due to many reasons. It is worth mentioning among them, for example, nonlinearity of the motion equations, variable coefficients or non lineal conditions on known or unknown borders. To solve these problems, different types of approximations which are combinations of analytical and numerical simulation methods are put into practice. The problem of plasma behavior in numerous varieties of a minimum-B magnetic trap where the plasma is heated under electron cyclotron resonance (ECR) conditions is the subject of numerical simulation studies. At present, the ECR minimum-B trap forms the principal part of the multicharge ion sources. 

There are different numerical methods to model plasmas. Depending of both temperature and concentration, these can be classified in three main groups: fluid models, kinetic models and hybrid models \cite{Birdsall}. The fluid models are the most simple way to describe the plasma from macroscopic quantities, which are used for the study of highly collisional plasmas where the mean free path is much smaller than size of plasma ($l_{mfp}\ll L$). The kinetic models are the most fundamental way to describe plasmas through the distribution function in phase-space for each particle specie; which are used for the study of weakly collisional ($l_{mfp}\sim L$) or collisionless plasmas ($l_{mfp}\gg L$) from the solution of the Boltzmann or Vlasov equation, respectively \cite{Verboncoeur}. For kinetic simulations there are different method to solve the Boltzmann or Vlasov equation, being the Particle-In-Cell (PIC) codes one the most popular \cite{Arber,Bruhwiler,Blumenfeld} The hybrid model combine both the fluid and kinetic models, treating some components of the system as a fluid, and others kinetically; which are used for the study of plasmas,  may use the PIC method for the kinetic treatment of some species, while other species (that are Maxwellian) are simulated with a fluid model.

In this work, a scheme of the relativistic Particle-in-Cell (PIC) code elaborated for an ECR plasma heating study in minimum-B traps is presented. For a PIC numerical simulation, the code is applied to an ECR plasma confined in a  minimum-B trap formed by two current coils generating a mirror magnetic configuration and a hexapole permanent magnetic bars to suppress the MHD instabilities. The plasma is maintained in a cylindrical chamber excited at $TE_{111}$ mode by $2.45$ $GHz$ microwave power. In the obtained magnetostatic field, the ECR conditions are fulfilled on a closed surface of ellipsoidal type. Initially, a Maxwellian homogeneous plasma from ionic temperature of $2$ $eV$ being during $81.62$ $ns$, that correspond to $200$ cycles of microwaves with an amplitude in the electric field of $1$ $kV/cm$  is heated. The electron population can be divided conditionally into a cold group of energies smaller than $0.2$ $keV$, a warm group whose energies are in a range of $3-10$ $keV$ and hot electrons whose energies are found higher than $10$ $keV$.

\end{abstract}

\keywords{Electron Cyclotron Resonance (ECR), minimum-B magnetic trap, Particle-In-Cell (PIC) method.}



\section{Introduction}

Actually, an electron cyclotron resonance ion source (ECRIS) is widely used for multicharge ions production because in the ECR conditions the plasma electrons can be energized by microwaves without any trouble \cite{geller1996electron}. In these systems, a minimum-B magnetic trap formed by the superposition of an axial magnetic field produced by two or more axisymmetric D.C. current coils and a radial magnetic field formed by a cusp multipole system is harnessed for plasma confining. Impossibility to apply the analytical methods for the ECRIS plasma study and practical inaccessibility of those plasmas for experimental diagnostics tools compel investigators to elaborate numerical methods so that they could obtain information about the behavior of both the plasma and individual particles in ECRIS \cite{valeriy_1,shirkov,valeriy_2}.

In this work, a numerical scheme developed as a relativistic full electromagnetic PIC code for plasma heating by microwaves under the ECR conditions in minimum-B traps is offered. In the PIC method chosen for our numerical simulations due to uncollisionality of the ECRIS plasma, the evolution of the plasma components is described by using superparticles (SP) \cite{yee,cern,umeda_1}. The simulations are carried out in three stages:

\begin{enumerate}
\item Simulation of the stationary microwave  field which is excited in the chamber through a port made in its cylindrical surface \cite{Alex}.
\item Fill the chamber with a homogeneous plasma ball in the instant $t=t_0$ and a Maxwellian distribution for electrons and ions velocities in such way that $\vec{J}(\vec{r},t=t_0)=0$ in order to not violate the continuity equation in the instant $t=t_0$.
\item Numerical study of the SP self-consistent  dynamics.
\end{enumerate}

The chamber of $4.54$ $cm$ in diameter and $10$ $cm$ long is excited at the electric transverse $TE_{111}$ mode by $2.45$ $GHz$ microwaves of $1$ $kV/cm$ in amplitude. The initial plasma is assumed maxwellian of a electronic temperature of $5.44 \times 10^{-4}$ $eV$ and a density of $2 \times 10^{10}$ $cm^{-3}$. 
 
The system can be described in the framework of the Vlasov-Maxwell equation and solved by using the particle-in-cell (PIC) method \cite{Lapenta,fitzpatrick2006computational,hockney1988computer}.\\
This paper is organized as follows: 

Section 2 describes the physical model and an electromagnetic PIC approach to the simulation of the plasma-microwave field interaction. In Section 3, the results concerning the plasma space distribution and the energy spectrum of electrons are depicted. 

\section{Physical system and numerical model} 
\label{section_2}

\subsection{Physical scheme model}

\begin{figure}[h]
	\centering
	\includegraphics[scale=0.35]{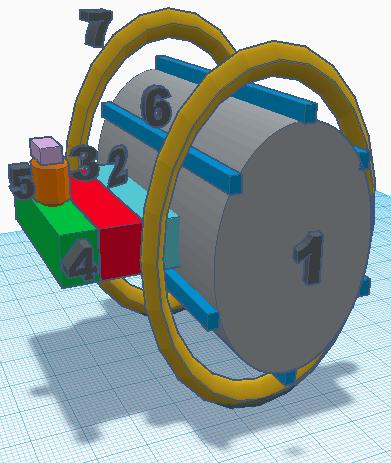}
	\caption{Physical system of an ECRIS: 1-Discharge chamber, 2-waveguide, 3-ferrite isolator, 4-waveguide, 5-magnetron, 6-magnetic hexapole, 7-coils.}
	\label{Phys_scheme}
\end{figure}

A physical model of the system under study is shown in Fig.\ref{Phys_scheme}. The ECR plasma in the non-magnetic metal chamber (1) is confined by the minimum-B magnetic field formed by two current coils (7) and  a cusp hexapole (6). The microwaves generated by a magnetron (5) are injected into a chamber (1) through waveguides (4) and (2).
\subsection{Numerical model}
\subsubsection{Electromagnetic particle-in-cell method}
\begin{figure}[h]
	\centering
	\includegraphics[scale=0.42]{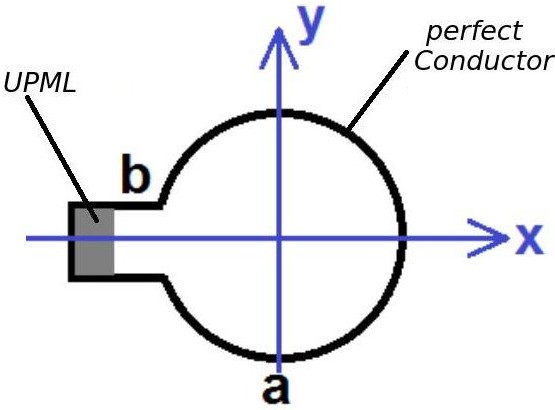}
	\caption{Waveguide-resonant cavity cross section.}
	\label{Cavity_PCE_UPML}
\end{figure}

The self-consistent simulation of the plasma-microwave interaction in our system is described in the framework of the Vlasov-Maxwell equation system:
\begin{eqnarray}
\frac{\partial f_{\alpha}}{\partial t}+\vec{v} \cdotp \nabla_{\vec{r}} f_{\alpha}
+\frac{q_\alpha}{m_\alpha}(\vec{E}+\vec{v} \times \vec{B}) \cdotp \nabla_{\vec{v}}\dfrac {f_{\alpha}}{\gamma}=0  \mbox{ }\mbox{ }\label{eq:Vlasov} 
\end{eqnarray}
where $\gamma=[1-(v_p/c)^2]^{-1/2}$ is the relativistic factor, $f_{\alpha}(\vec{r},\vec{v},t)$ is a six-dimentional phase space distribution for a specie $\alpha$, electron or ion in our case, whose charge and mass are $q_\alpha$ and $m_\alpha$, respectively.\\ $\vec{E}$ and $\vec{B}$ are the total electric and magnetic fields, respectively. In our system, $\vec{E}=\vec{E}^{sc}$ correspond to the self-consistent electric field, which is the superposition of the microwave field $E^{hf}$ and the plasma self-generated electric field $E^{sg}$. \\ $\vec{B}=\vec{B}^{s}+\vec{B}^{sc}$, where $\vec{B}^{s}$ is the magnetostatic field of the minimum-B magnetic trap is calculated through a procedure described in \cite{MaoOswaldo2016}. $\vec{B}^{sc}$ is the self-consistent magnetic field produced due to the plasma particle motions. which has a similar meaning of the electric field component previously described.
The evolution of the $\vec{E}^{sc}$ and $\vec{B}^{sc}$ are given by the following Maxwell equations: 
\begin{eqnarray}
 \vec{\nabla}\times\vec{E}^{sc}=-\frac{\partial\vec{B}^{sc}}{\partial t}, & &   \vec{\nabla}\times\vec{B}^{sc}=\mu_{0}\vec{J}-\varepsilon_{0}\frac{\partial\vec{E}^{sc}}{\partial t} \label{eq:rot_E_B}
\end{eqnarray}
The charge density and current density are determined by the expressions:
\begin{eqnarray}
\rho(\vec{r},t)=\sum_{\alpha} q_\alpha\int \gamma f_{\alpha} d\vec{v}, & \mbox,  \vec{J}(\vec{r},t)=\sum_{\alpha} q_\alpha\int  \gamma \vec{v}f_{\alpha} d\vec{v} \label{eq:rho_J}
\end{eqnarray}
respectively.\\
In PIC simulations, the distribution functions of each specie is given by the superposition of SPs which are the clouds of real particles \cite{Lapenta}:\\
\begin{equation}
 f_{\alpha}(\vec{r},\vec{v},t)=\sum_{p} f_p(\vec{r},\vec{v},t)
\end{equation}
In our simulation the state of Np-particle ensemble at time t  is specified by the exact  one-SP distribution function:
\begin{equation}
 f_p(\vec{r},\vec{v},t)=N_p S_{\vec{r}}\big(\vec{r}-\vec{r}_p(t)\big)\delta \big(\vec{v}-\vec{v}_p(t)\big)
\end{equation}
Here $N_p$ is the number of real particles that form a SP; $\delta \big(\vec{v}-\vec{v}_p(t)\big)$ is the Dirac delta function for the velocities distribution, being $\vec{v}_p$ the velocity distribution. The spatial shape factor is given by,
\begin{equation}
S_{\vec{r}}\big(\vec{r}-\vec{r}_p(t)\big)=\prod_{j=1}^3 b_0\Bigg(\frac{x_j-x_{jp}(t)}{\Delta x_j} \Bigg)
\end{equation}
where the $j$-index refer to the $x$, $y$ and $z$ coordinates, $\vec{r}_p$ is the SP position; $\Delta x_j$ is the spatial step in $j$ direction, and
\begin{eqnarray} 
 b_0(\xi)= 
   \begin{cases} 
      1     & \mbox{if }  \arrowvert \xi \arrowvert<1/2   \\
      0     & \mbox{otherwise}
   \end{cases}
\end{eqnarray}
corresponds to the first b-spline.\\
The SP dynamics determined by the relativistic Newton-Lorentz equation is solved numerically through the Boris leapfrog procedure \cite{taflove2000computational}:
\begin{eqnarray}
 \frac{d\vec{r}_p}{dt}=\vec{v}_p,& \mbox{ } & \frac{d(\gamma m_s \vec{v}_p)}{dt}=q_s (\vec{E}_p+\vec{v_p}\times \vec{B}_p) \label{Motion_Eq}
\end{eqnarray}
where $\gamma=[1-(v_p/c)^2]^{-1/2}$ is the relativistic factor, $q_s=N_p q_\alpha$ and $m_s=N_p m_\alpha$ are the charge and
mass of a \textit{SP}. Such equations are very similar to those describing the motion of the real particles but the fields $\vec{E}_p$ and $\vec{B}_p$ are calculated as an average over the SP to preserve the moments of the Vlasov equation. These fields can be calculated from the values calculated in gridpoints, $\vec{E}_g$ and $\vec{B}_g$, as
\begin{equation}
 \vec{E}_p=\sum_g \vec{E}_g W(\vec{r}_g-\vec{r}_p)d\vec{r}	
\end{equation}
and
\begin{equation}
 \vec{B}_p=\sum_g \vec{B}_g W(\vec{r}_g-\vec{r}_p)d\vec{r}	
\end{equation}
respectively. The interpolation function $W$ depends on the choosen spatial shape factor. For the present case  
\begin{equation}
W\big(\vec{r}_g-\vec{r}_p\big)=\prod_{j=1}^3 b_1\Bigg(\frac{x_{jg}-x_{jp}(t)}{\Delta x_{j}} \Bigg)
\end{equation}
where 
\begin{eqnarray} 
 b_1(\xi)= 
   \begin{cases} 
      1-\arrowvert \ \xi \arrowvert  & \mbox{if }  \arrowvert \xi\arrowvert<1   \\
      0        & \mbox{otherwise}
   \end{cases}
\end{eqnarray}
corresponds to the second b-spline.\\
In our numerical model, it is supposed that the considered chamber is made of a perfect electric conductor (PEC) and in order to avoid nonphysical reflections at the microwave port, the Uniaxial Perfectly Matched Layer (UPML) method is used (See Fig.\ref{Cavity_PCE_UPML}).\\
The Maxwell rotational equations (\ref{eq:rot_E_B}) at UPML conditions are expressed in a phasor-like form \cite{taflove2000computational}:

\begin{equation}
\centering
\vec{\nabla}\times \vec{H}^{sc*}=\vec{J}^*+j\omega \varepsilon_0 \overline{\overline S} \vec{E}^{sc*} \label{eq: Ampere_Maxwell}
\end{equation} 
and
\begin{equation}
\centering
\vec{\nabla}\times \vec{E}^{sc*}=-j\omega \mu_0 \overline{\overline S} \vec{H}^{sc*}\label{eq: Faraday}
\end{equation}
where $\overline{\overline S}$ is a diagonal tensor defined by
\[
\overline{\overline S}=
  \begin{bmatrix}
    s_x^{-1}s_ys_z & 0 & 0 \\
    0 & s_xs_y^{-1}s_z & 0 \\
    0 & 0 & s_xs_ys_z^{-1}
  \end{bmatrix},
\]
where:
\begin{equation}
s_x = k_x + \frac{\sigma_x}{j\omega\varepsilon_0} \  \ ; \  \  s_y = k_y + \frac{\sigma_y}{j\omega\varepsilon_0}\  \ ; \  \  s_z = k_z + \frac{\sigma_z} {j\omega\varepsilon_0}
\end{equation}
In our simulations $k_x$, $k_y$ y $k_z$ are chosen equal to the unity and the electrical conductivities $\sigma_x,\sigma_y,\sigma_z$ are chosen equal to zero in all space except the microwave port where $\sigma_x=(x/d)^3\sigma_x^{max}$, $d=10\Delta x$ is the width of the UPML with the spatial step $\Delta x$ in $x$ direction, $x$ is the distance from the border  towards the interior of the UPML and $\sigma_x^{max}=0.8(m+1)/(\Delta x \sqrt{\mu_0/\varepsilon_0})$ with $m=3$, because 10 cells are used in the UPML zone.\\
Starting from the law of Ampere-Maxwell (\ref{eq: Ampere_Maxwell}), defining the constitutive relations as:
\begin{equation}
D_x^{sc}=\varepsilon_0 \frac{s_z}{s_x} E_x^{sc} \ , \ D_y=\varepsilon_0 \frac{s_x}{s_y} E_y^{sc}\ , \ D_z^{sc}=\varepsilon_0 \frac{s_y}{s_z} E_z^{sc}\label{eq_4_dx}
\end{equation}
and applying a Fourier inverse transform, a set of equivalent equations in the time domain are obtained. In a finite difference form for the $x$ component, it can be written:
{\small{
\begin{eqnarray}
E_x^{n+1}(i+1/2,j,k)=\frac{\left[1-\frac{\sigma_z(k)\Delta t}{2\varepsilon_0 k_z(k)}\right]}{\left[1+\frac{\sigma_z(k)\Delta t}{2 \varepsilon_0 k_z(k)}\right]} E_x^n (i+1/2,j,k) \nonumber \\
+ \frac{k_x(i+1/2)}{\varepsilon_0 k_z(k)}\frac{\left[1-\frac{\sigma_x(i+1/2)\Delta t}{2\varepsilon_0  k_x(i+1/2)}\right]}{\left[1+\frac{\sigma_z(k)\Delta t}{2 \varepsilon_0 k_z(k)}\right]}D_x^{n+1}(i+1/2,j,k) \nonumber \\ -\frac{k_x(i+1/2)}{\varepsilon_0 k_z(k)}\frac{\left[1-\frac{\sigma_x(i+1/2)\Delta t}{2\varepsilon_0  k_x(i+1/2)}\right]}{\left[1+\frac{\sigma_z(k)\Delta t}{2 \varepsilon_0 k_z(k)}\right]}D_x^{n}(i+1/2,j,k)
\end{eqnarray}}}
and
{\small{\begin{equation}
 D_x^{n+1}(i+1/2,j,k) = \frac{\left[1-\frac{\sigma_y(j)\Delta t}{2\varepsilon_0 k_y(i)}\right]}{\left[1+\frac{\sigma_y (j)\Delta t}{2\varepsilon_0 k_y(j)}\right]}D_x^n(i+1/2,j,k) + \frac{\Delta t}{k_y(j)\left[1+\frac{\sigma_y(j)\Delta t}{2\varepsilon_0 k_y(j)}\right]}\cdot \nonumber \\
 \end{equation}
 \begin{equation}
 \left[\frac{H_z^{n+1/2}(i+1/2,j+1/2,k)-H_z^{n+1/2}(i+1/2,j-1/2,k)}{\Delta y} \right. \nonumber
  \end{equation}
  \begin{equation}
- \left. \frac{H_y^{n+1/2}(i+1/2,j,k+1/2)-H_y^{n+1/2}(i+1/2,j,k-1/2)}{\Delta z}\right] \nonumber
\end{equation}
\begin{equation}
 -J^{n+1/2}(i+1/2,j,k)    \label{Dx}
\end{equation}
}}

Where $n$ denote the index for the time $t_n=n\Delta t$. For simplicity, in the previous equations the superscript $"sc"$ has been omitted.\\ 
Similarly, from the law of Faraday (\ref{eq: Faraday}) and the constitutive relations:
\begin{equation}
B_x^{sc}=\mu_0 \frac{s_z}{s_x} H_x^{sc} \ , \ B_y^{sc}=\mu_0 \frac{s_x}{s_y} H_y^{sc}\ , \ B_z^{sc}=\mu_0 \frac{s_y}{s_z} H_z^{sc}, \label{B_components}
\end{equation}
the equations for the evolution of the rectangular components of the magnetic field are obtained.\\

\begin{figure}[h!]
	\centering
	\includegraphics[scale=0.15]{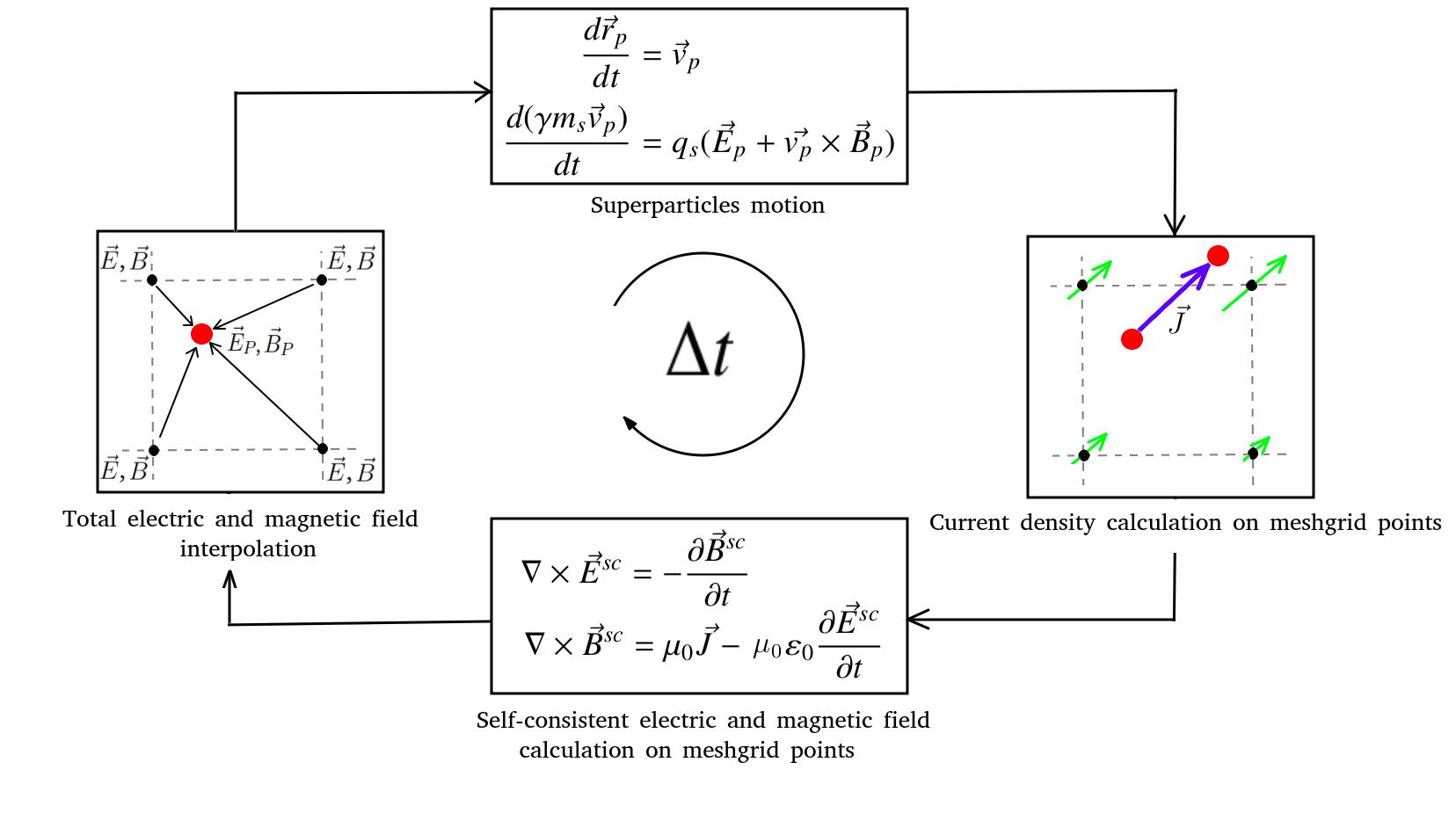}
	\caption{\label{PIC}  Computational PIC cycle.}
\end{figure}

In Fig.\ref{PIC}, the computational cycle steps are shown:

\begin{enumerate}
\item calculation of the current densities in the mesh nodes
beginning with the SP positions and velocities data; 
\item computation of the total fields in the mesh nodes and their velocities data by using the charge conservation method \cite{fitzpatrick2006computational}; 
\item calculation of the total electric and magnetic fields in the mesh nodes,
\item calculation of the new positions and velocities of the SPs through integration of their motion equations.
\end{enumerate}

The electric and magnetic fields calculated in the mesh nodes and then interpolated to the SP positions make it possible to determine the forces acting on particles that predetermines the SP positions and their impulses on the next time step (See Fig.\ref{PIC}).

\section{Results}
\label{section_7}

The process of propagation of $290$ $kW$ microwave power along a $TE_{10}$ rectangular waveguide and its penetration into the chamber through a port made in its cylindrical surface is numerically simulated. In the stationary regime, a microwave field of $1$ $kV/cm$ tension is set. The chamber quality factor is of $50$.

\begin{figure}[h!]
	\centering
	\includegraphics[scale=0.34]{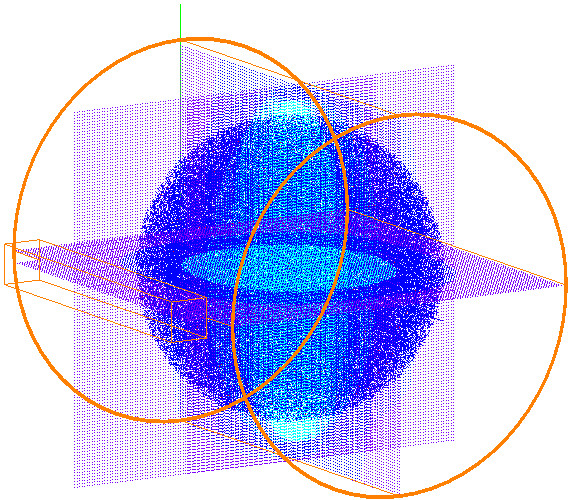}
	\caption{\label{screenshot} Screenshot of the simulation for the instant when the chamber is filled with
		the plasma.}
\end{figure}

\begin{figure}[h!]
	\centering
	\includegraphics[scale=0.60]{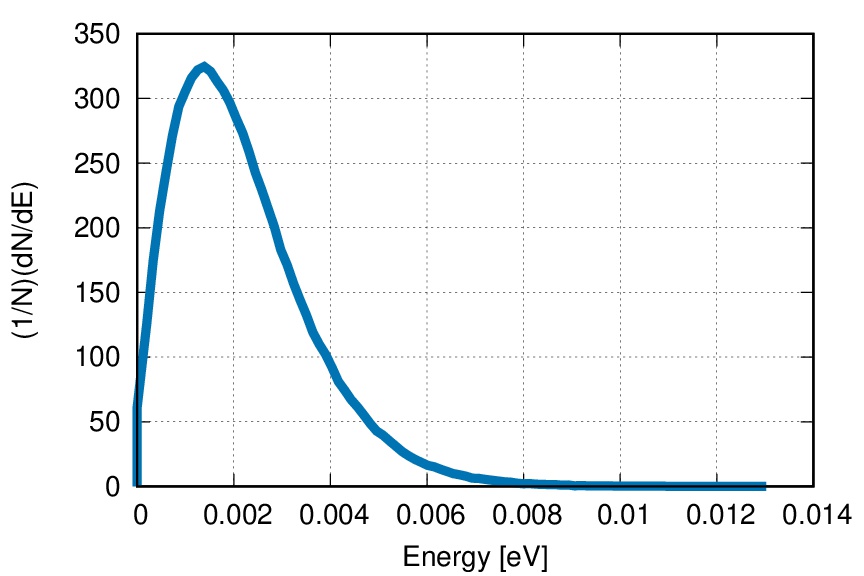}
	\caption{\label{distrib_initial} Energy spectrum of electrons for the instant when the chamber is filled
		with the plasma.}
\end{figure}

The simulations are fulfilled on a rectangular $3D$ mesh at the spatial steps $\Delta x$ $=$ $\Delta y$ $=$ $0.07$ $cm$, $\Delta z$ $=$ $0.2$ $cm$ at a time step $\Delta t$ $=$ $2.07$ $ps$, that are chosen in accordance with the Courant stability condition. The initial plasma configuration presents a sphere of $2.27$ $cm$ in radius spaced in the center of the chamber.The energy distribution function is maxwellian at a maximum of $2$ $meV$ (See Figs.\ref{screenshot} and \ref{distrib_initial}).

The total number of $1.5 \times 10^{12}$ particles is divided into $3 \times 10^{6}$ superparticles that reciprocally interact through their electric and magnetic fields.
In the course of simulations, the plasma particles are steeply redistributed in the chamber volume, but their total number remains fixed.

\begin{figure}[h]
	\centering
	\includegraphics[scale=0.19]{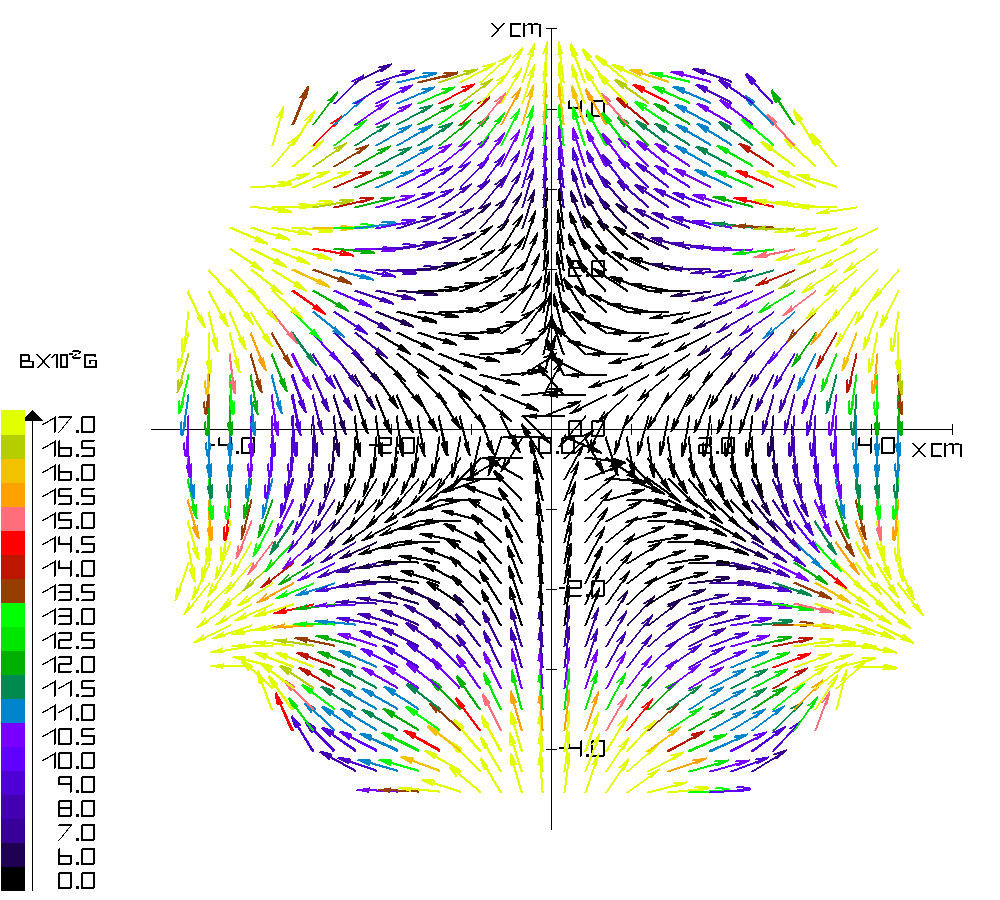}
	\caption{\label{transverse_section} Magnetostatic field lines in the transverse plane z= 5 cm.}
\end{figure}

\begin{figure}[h]
	\centering
	\includegraphics[scale=0.20]{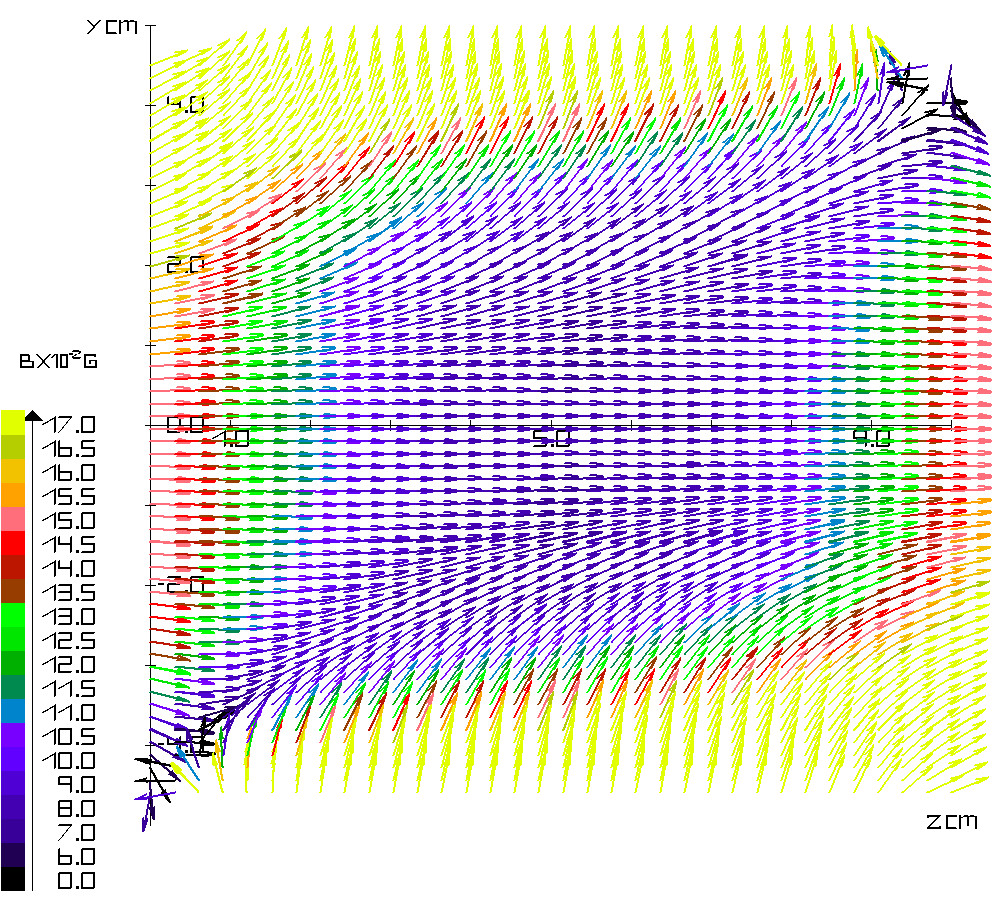}
	\caption{\label{longitudinal_section} Magnetostatic field lines in a longitudinal plane x = 0.}
\end{figure}

The calculated vector image of the magnetostatic field in the transverse plane is shown in Fig.\ref{transverse_section}, and in the longitudinal plane in Fig.\ref{longitudinal_section}. In Fig.\ref{transverse_section}, we can see that the magnetic field generated by the hexapolar system, has the typical configuration of a minimum-B magnetic trap which contributes to the radial confinement of the plasma. On the other hand, in Fig.\ref{longitudinal_section}, one can see the magnetistatic field lines in a longitudinal plane.

\begin{figure}[h!]
	\centering
	\includegraphics[scale=0.71]{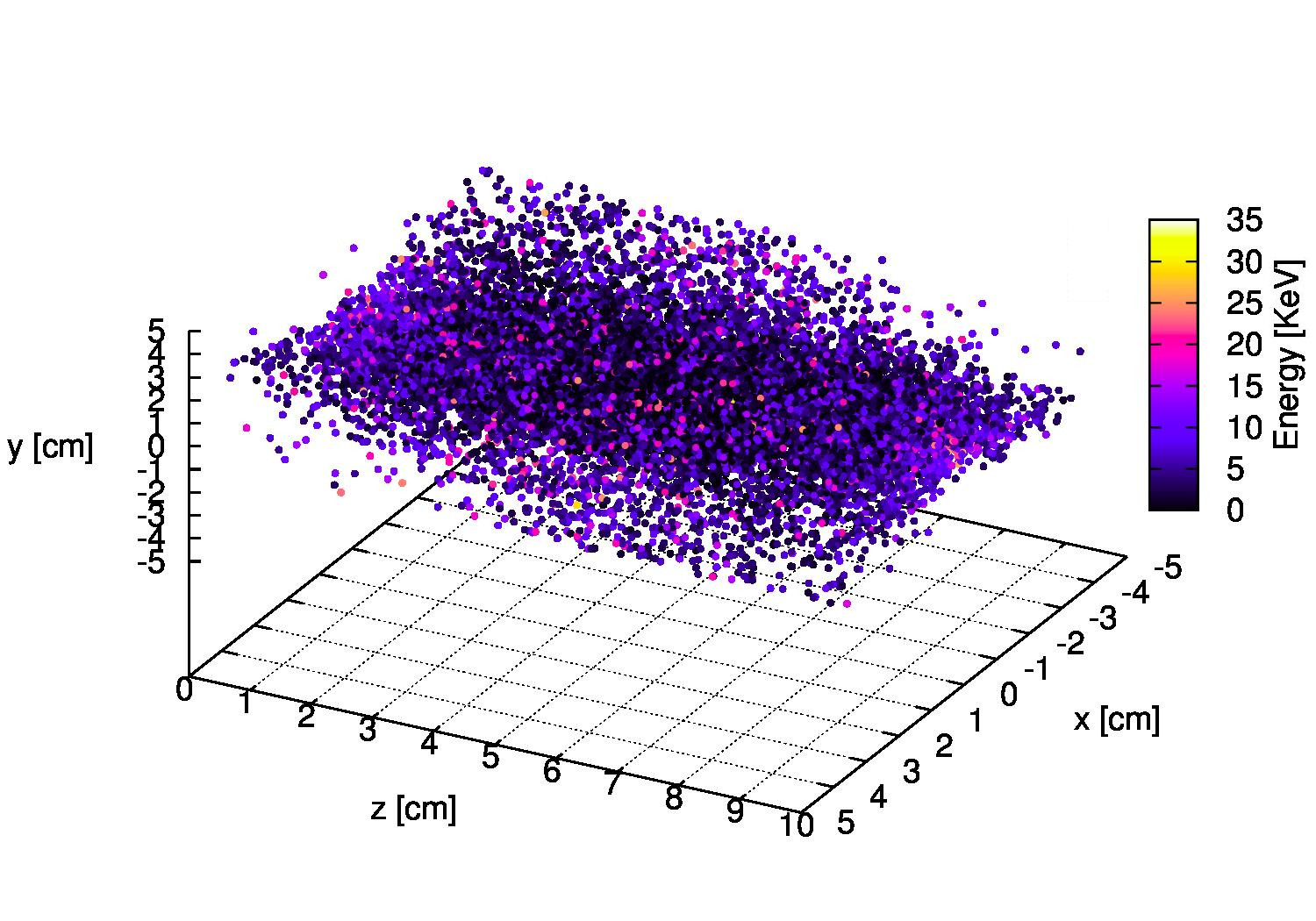}
	\caption{\label{dist_3d} Spatial distribution for electrons and its energies after 200 cycles of the microwave field from the instant when the chamber is filled with the plasma.}
\end{figure}

\begin{figure}[h!]
	\centering
	\includegraphics[scale=0.17]{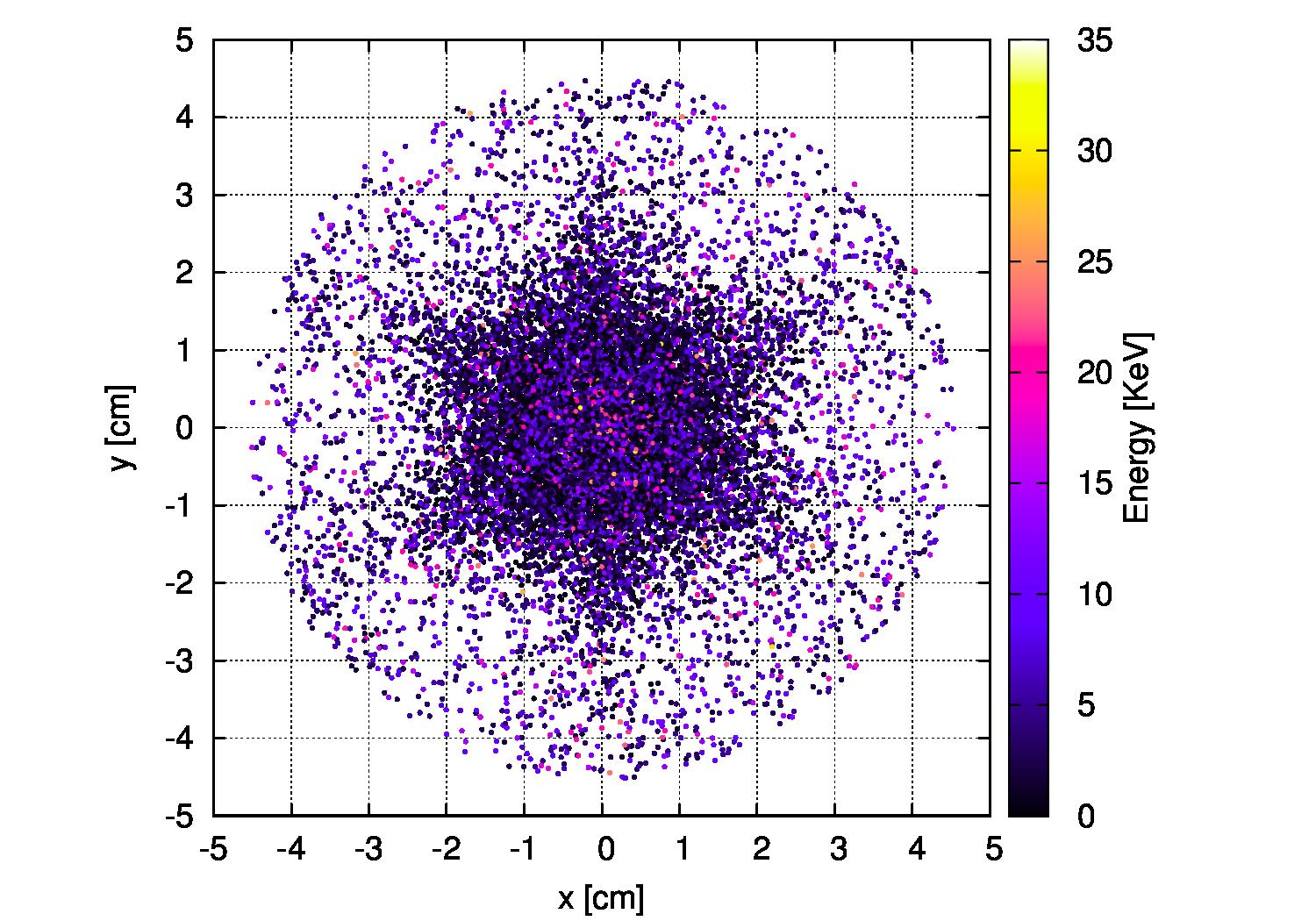}
	\caption{\label{dist_2d} Spatial distribution of electrons with different energies.}
\end{figure}

In Fig.\ref{dist_3d} and Fig.\ref{dist_2d}, we show the results concerning the energy spatial distributions for electrons after 200 cycles of microwaves, where one can see that the core plasma density exceeds by far the density of the corona region. In general, the plasma is accumulated in the trap axis vicinity and symmetrically to the center of the trap.

\begin{figure}[h!]
	\centering
	\includegraphics[scale=0.55]{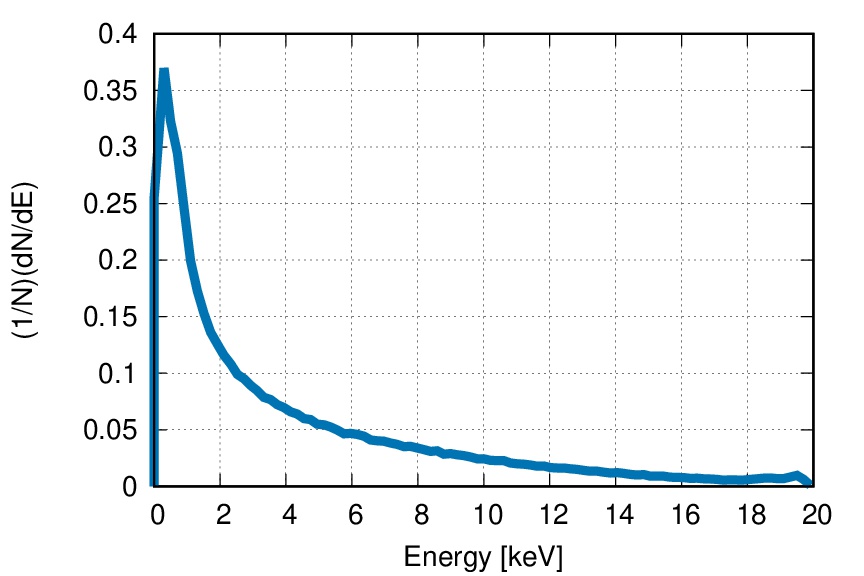}
	\caption{\label{final_dist_1} Distribution of electrons after 55 cycles of microwaves with different energies in the central transverse plane of the chamber.}
\end{figure}

\begin{figure}[h!]
	\centering
	\includegraphics[scale=0.55]{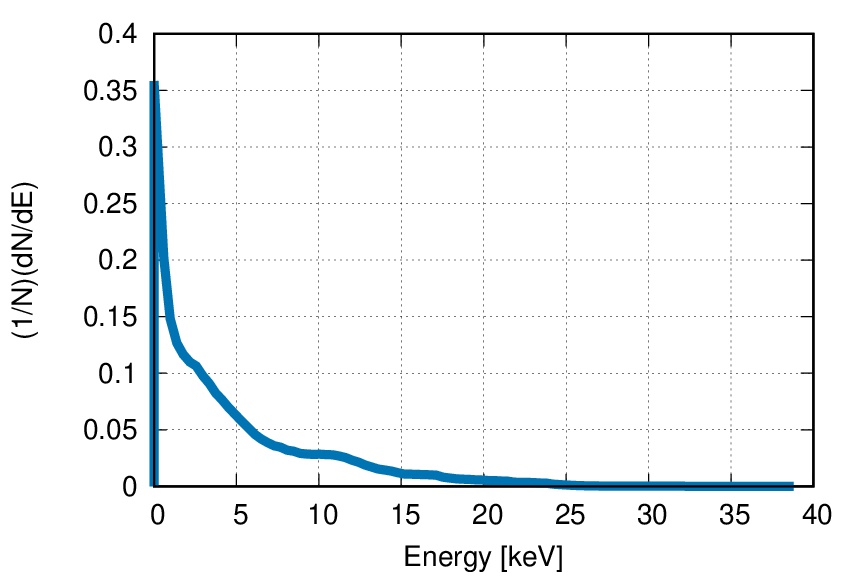}
	\caption{\label{final_dist_2} Distribution of electrons after 200 cycles of microwaves with different energies in the central transverse plane of the chamber.}
\end{figure}

The data shown in Fig.\ref{final_dist_1} and Fig.\ref{final_dist_2} evidence that the electron population can be divided into 3 groups: the cold electrons with energies less than $0.2$ $keV$, the electrons with intermediate energy between $0.2$ and $3$ $keV$, the warm electrons whose energies range from $3$ to $10$ $keV$ and the hot electrons with energies higher than $10$ $keV$. An estimation shows that the intermedite and warm electrons constitute 90.3\% with reference to the total electron population, the cold electrons contribution is 9.5\% and the hot electron fraction is 0.2\%.

\newpage

\section{Conclusions}
\label{section_8}

A numerical scheme for simulating the behavior of collisionless plasmas confined in the ECR minimum-B magnetic traps which include the simulation of chamber excitation is elaborated.
This scheme implemented for determining the characteristics of ECR plasma heating in a minimum-B magnetic trap shows that the electron component is formed by three groups: cold, warm and hot. Appearing these groups can be attributed to different modes of the microwave electron interaction. This point will be elucidated through some more long-run simulations. The hot group, which is of interest to ECRIS is localized predominantly in the trap center.

\section*{Acknowledgments}

The work is supported by the Universidad Industrial de Santander under the program of mobility VIE and for the GridUIS-2 testbed of SC3UIS. One of the authors (Alex Estupi\~n\'an) would like to thank the Universidad Aut\'onoma de Bucaramanga.

\end{document}